\documentclass[12pt]{article}
\usepackage{amsmath,amsthm,amssymb}
\usepackage{mathtools}
\usepackage{geometry}
\usepackage{booktabs}
\usepackage{graphicx}
\usepackage{hyperref}
\usepackage[numbers,sort&compress]{natbib}
\usepackage{microtype}
\usepackage{setspace}
\usepackage{enumitem}
\usepackage{xcolor}
\usepackage{float}
\theoremstyle{plain}
\newtheorem{theorem}{Theorem}[section]

\newtheorem{proposition}[theorem]{Proposition}

\theoremstyle{definition}
\newtheorem{definition}[theorem]{Definition}
\newtheorem{assumption}[theorem]{Assumption}
\newtheorem{axiom}{Axiom}
\theoremstyle{remark}
\newtheorem{remark}[theorem]{Remark}

\newcommand{\calS}{\mathcal{S}}
\newcommand{\calF}{\mathcal{F}}

\newcommand{\PP}{\mathbb{P}}
\newcommand{\EE}{\mathbb{E}}

\newcommand{\Disc}{\mathrm{Disc}}
\newcommand{\Sev}{\mathrm{Sev}}

\newcommand{\dAUC}{\Delta\mathrm{AUC}}

\newcommand{\fail}{\mathrm{Fail}}
\newcommand{\death}{\mathrm{Death}}

\begin{document}

\begin{flushleft}
{\Large\textbf{A statistical framework for identifying subgroup vulnerability to predictive multiplicity in clinical AI}}\\
\textit{Short title: Subgroup vulnerability to predictive multiplicity}\\[1em]
Enock Adu Bonsu\textsuperscript{1*}\\
\bigskip
\textbf{1} Department of Epidemiology and Biostatistics, Mel and Enid Zuckerman College of Public Health, University of Arizona, Tucson, Arizona, United States of America\\
\bigskip
* Corresponding author: enocka@arizona.edu
\end{flushleft}

\newpage

\section*{Abstract}
Artificial intelligence (AI) models are increasingly used to predict patient outcomes, yet models trained on the same data can disagree about patient risk, with disagreement potentially concentrated in clinically important subgroups. We propose $V(S)$, a statistically grounded vulnerability index that combines an observable lower-bound witness of model disagreement with clinical severity, and develop inference and multiplicity-adjustment procedures for auditing prespecified subgroups. We applied the framework to two large critical-care cohorts, MIMIC-IV ($n=65{,}078$) for model development and eICU-CRD ($n=188{,}230$ admissions, 208 hospitals) for external validation, comparing a random forest with logistic regression across 158 prespecified subgroups. The two primary models did not both satisfy the prespecified $\epsilon=0.02$ Rashomon-set tolerance: the logistic-regression AUC was 0.0488 below the best candidate-model AUC. Accordingly, the RF--LR discrimination gap is interpreted as disagreement between two specific models rather than as a formally guaranteed lower bound on the full Rashomon set. Nine subgroups had discrimination gaps distinguishable from a prespecified clinical floor. The age $\geq80$ and cardiac subgroup had the largest point estimate of $V(S)$ ($0.307$), but was underpowered and did not meet the full high-priority decision rule. The univariate cardiac subgroup ($\hat V(S)=0.193$, 95\% confidence interval [0.163, 0.223]) was the only statistically distinguishable subgroup with adequate power. Post hoc analyses identified lactate as an important variable for both models, but did not establish a causal or mechanistic explanation for the model disagreement. Four simulation studies quantified the operating characteristics of the proposed procedures, including inflated small-sample detection rates and imperfect Wald-interval coverage. The framework provides a reproducible approach for ranking subgroup vulnerability to model disagreement while explicitly separating exploratory signals from adequately supported findings.

\section*{Author Summary}
Different artificial intelligence models can make different predictions for the same patients even when they perform similarly when evaluated across an entire hospital population. This matters because the choice of model can affect which patients are identified as high risk. I developed a statistical framework that helps identify patient subgroups in which this disagreement may be especially important. The framework combines the size of the disagreement between models with the seriousness of outcomes in the subgroup and accounts for the fact that many subgroups may be examined at once. I evaluated the approach using critical-care data from more than 250,000 admissions across two large databases. The cardiac-diagnosis subgroup showed a reproducible signal of disagreement between a random forest and logistic regression, although only the broad cardiac subgroup had enough observations to support a statistically distinguishable result. A smaller age-and-cardiac subgroup had a high point estimate but was too uncertain to be classified as high priority. Simulations also showed that inference is less reliable for very small subgroups. These findings illustrate why subgroup auditing should report uncertainty and statistical power rather than relying on point estimates alone.

\newpage
\section{Introduction}

Artificial intelligence models are being adopted across clinical medicine for risk prediction, triage, resource allocation, and treatment support \citep{bonsu2024leveraging,obermeyer2019dissecting}. A recurring concern is that models with good aggregate performance can behave differently for specific patient subgroups. Moreover, multiple models can achieve similar overall performance while producing materially different predictions, a phenomenon described as predictive multiplicity \citep{breiman2001two,fisher2019all,marx2020predictive,black2022model}. When disagreement is concentrated in a clinically important subgroup, model selection can therefore have consequences that are not visible from population-level discrimination alone.

Existing subgroup-auditing approaches often rely on direct stratification of performance metrics or post hoc explainability. Stratification can reveal subgroup differences but does not, by itself, provide a principled way to rank many subgroups while accounting for uncertainty and multiplicity. Explainability tools such as SHAP can characterize feature contributions but do not directly quantify disagreement between alternative prediction models or establish which subgroups warrant additional validation.

We develop $V(S)$ as a framework for prospective statistical auditing of prespecified patient subgroups. The framework combines a discrimination-disagreement component with a clinical-severity component, provides confidence intervals and a composite decision rule, and incorporates multiplicity adjustment across subgroups. The theoretical motivation comes from predictive multiplicity: if two models belong to a class of models that are comparably good overall, their subgroup-specific AUC difference is necessarily no larger than the maximum disagreement over that class. This relationship is conditional on membership in the comparison class and is therefore explicitly verified rather than assumed.

We apply the framework to in-hospital mortality prediction using a random forest (RF) and logistic regression (LR) trained on MIMIC-IV and evaluated with eICU-CRD as an external critical-care cohort. The application is designed to illustrate both the utility and the limitations of subgroup auditing. In particular, we report subgroup power, multiplicity-adjusted decisions, robustness across additional model pairs and hospital sites, and simulation-based evidence about small-sample inference.

\section{Materials and Methods}

\subsection{Study Design and Data Sources}
This methodological study develops and evaluates a framework for auditing subgroup vulnerability to disagreement among clinical prediction models. Development used MIMIC-IV v3.1 \citep{johnson2023mimiciv}; external validation used eICU-CRD v2.0 \citep{pollard2018eicu}. After the stated inclusion criteria, MIMIC-IV contributed 65,078 ICU admissions and eICU-CRD contributed 188,230 admissions across 208 hospitals. The outcome was in-hospital mortality. Fifteen prespecified features were used: age, sex, race/ethnicity, heart rate, systolic/diastolic/mean blood pressure, temperature, SpO$_2$, creatinine, hemoglobin, white blood cell count, lactate, Charlson comorbidity index, and admission diagnosis category.

\begin{figure}[H]
\centering
\includegraphics[width=0.85\textwidth]{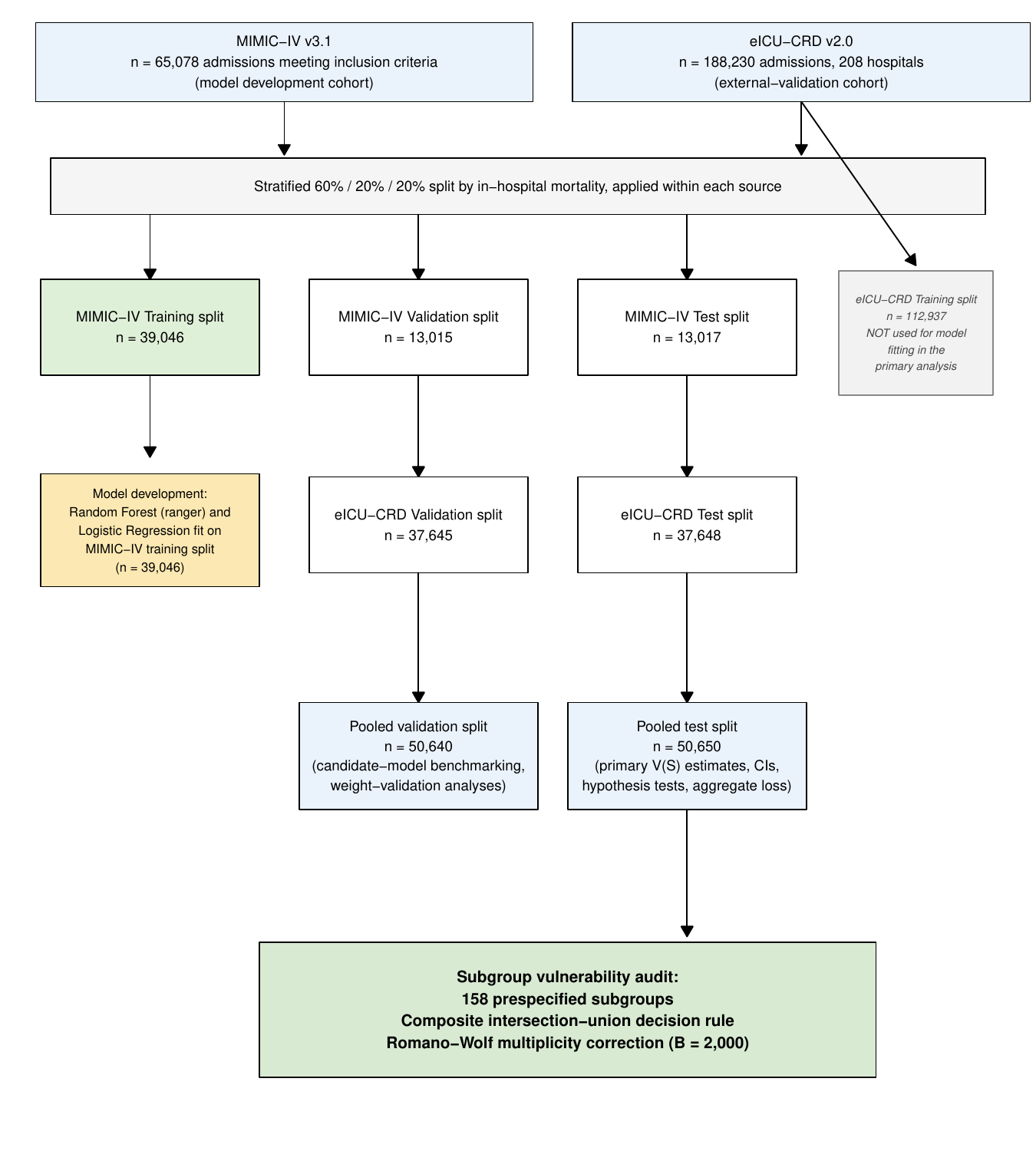}
\caption{\textbf{Study data-flow diagram.} MIMIC-IV and eICU-CRD admissions meeting inclusion criteria are each split 60\%/20\%/20\% (train/validation/test) stratified by in-hospital mortality. Only the MIMIC-IV training split is used for model fitting; the eICU-CRD training partition is not used in the primary analysis. A small number of validation/test rows (20 and 15, respectively) are subsequently excluded because they contain a categorical predictor level not present in the MIMIC-IV training split, producing an undefined model prediction. Validation and test partitions from both sources are pooled for candidate-model benchmarking and the primary subgroup vulnerability audit, respectively.}
\label{fig:flow}
\end{figure}

Data were split 60\%/20\%/20\% within each source, stratified by outcome. Models were trained on the MIMIC-IV training split ($n=39{,}046$). The pooled validation split ($n=50{,}640$) was used for candidate-model benchmarking and weight-validation analyses, while the pooled test split ($n=50{,}650$) was reserved for the primary subgroup estimates, confidence intervals, hypothesis tests, and aggregate loss calculations. Figure~\ref{fig:flow} summarizes the flow of admissions from both sources through model development and the pooled evaluation splits.

Missing values among continuous predictors were imputed using the median value computed on the training split only, applied unchanged to the validation and test splits; a corresponding missingness indicator was retained for each imputed predictor. Missingness was concentrated in lactate and temperature.

\subsubsection{Ethics Statement}
This study did not involve identifiable human subjects data and did not require Institutional Review Board approval. MIMIC-IV and eICU-CRD are publicly available, de-identified critical-care datasets accessed under the applicable PhysioNet Data Use Agreements, following completion of the required human-subjects research training (CITI Program) by the author. MIMIC-IV and eICU-CRD were collected and de-identified under source-institution protocols described in the original database publications \citep{johnson2023mimiciv,pollard2018eicu}. No individually identifiable information was used in this secondary analysis.

\subsubsection{Patient and Public Involvement}
No patients or members of the public were involved in the design, conduct, reporting, or dissemination of this study. This is a secondary methodological analysis of de-identified, publicly available critical-care registry data (MIMIC-IV and eICU-CRD); direct patient or public involvement was not applicable given the retrospective, registry-based design and the absence of direct contact with patients or their records at the individual level.

\subsubsection{Data and Code Availability}

MIMIC-IV v3.1 and eICU-CRD v2.0 are publicly available through PhysioNet to credentialed users who complete the required human-subjects research training and data use agreement. The author completed the required credentialing and accessed both datasets under the applicable PhysioNet Data Use Agreements. Patient-level data cannot be redistributed by the author; researchers must obtain independent credentialing and authorization through PhysioNet to access the source data. All analysis code, cohort-derivation logic, and extraction queries used in this study are publicly available through Zenodo \citep{bonsu2026code} and are maintained at \url{https://github.com/Enockadubonsu/vulnerability-index-}.

\subsubsection{Cross-Dataset Race/Ethnicity Harmonization}
MIMIC-IV and eICU-CRD use different race/ethnicity vocabularies. MIMIC-IV granular labels were rolled up to parent categories, corresponding eICU labels were mapped to the shared vocabulary, and undocumented or unknown categories were retained in a canonical Unknown/Other category. The complete mapping used for analysis should be supplied with the code repository so that this harmonization is reproducible.

\subsection{Mathematical Framework}
\subsubsection{Theoretical Foundation via Predictive Multiplicity}
Let $\calS$ denote patient subgroups and let $\calF_\epsilon$ denote the set of models whose overall-population discrimination is within $\epsilon$ of the best candidate-model AUC on the validation data. Let $\pi_S=\PP(\death\mid S)$.

\begin{definition}[Predictive Multiplicity]
For subgroup $S$,
\begin{equation}
M(S)=\sup_{f,g\in\calF_\epsilon}|\mathrm{AUC}_S(f)-\mathrm{AUC}_S(g)|.
\end{equation}
This quantity represents the largest subgroup-specific discrimination disagreement among models satisfying the prespecified overall-performance criterion.
\end{definition}

\begin{assumption}[Failure Mechanism]
$\fail(S)\iff M(S)>\gamma$.
\end{assumption}

\begin{assumption}[Failure Consequence]
$\PP(\death\mid\fail,S)=\pi_S$.
\end{assumption}

\begin{definition}[Theoretical Vulnerability Index]
\begin{equation}
V_{\mathrm{theor}}(S)=\PP(\fail\mid S)\PP(\death\mid\fail,S).
\end{equation}
\end{definition}

\begin{remark}[Interpretation of the theoretical quantity]
Because $\fail(S)$ is defined deterministically at the subgroup level, the preceding quantity is not itself an operational estimator with a continuously varying failure probability. It is used only as conceptual motivation for combining disagreement and clinical consequence; the operational index below is defined independently.
\end{remark}

\begin{proposition}[Conditional Lower-Bound Witness]
If $f,g\in\calF_\epsilon$, then
\begin{equation}
|\mathrm{AUC}_S(f)-\mathrm{AUC}_S(g)|\leq M(S).
\end{equation}
Thus an observed pairwise disagreement is a lower-bound witness for the supremum only when both models satisfy the defining $\calF_\epsilon$ criterion.
\end{proposition}

\begin{remark}[Application to RF and LR]
The RF--LR pair does not satisfy this condition at $\epsilon=0.02$ because LR is 0.0488 below the best validation AUC. The RF--LR difference is consequently used in the empirical application as a model-specific disagreement measure rather than as a formally guaranteed lower bound on $M(S)$.
\end{remark}

\subsubsection{Axiomatic Motivation for the Functional Form}
We require the vulnerability functional $\mathcal V(D,M)$ to be nondecreasing in both disagreement $D$ and severity $M$, and to allow their joint contribution to increase vulnerability. We therefore impose the following properties.

\begin{axiom}[Monotonicity]
$\mathcal V(D,M)$ is nondecreasing in each argument.
\end{axiom}

\begin{axiom}[Supermodularity \citep{topkis1998supermodularity}]
\label{ax:super}
For $D_1\geq D_2$ and $M_1\geq M_2$,
\begin{equation}
\mathcal V(D_1,M_1)-\mathcal V(D_2,M_1)\geq \mathcal V(D_1,M_2)-\mathcal V(D_2,M_2).
\end{equation}
\end{axiom}

\begin{axiom}[Smoothness]
$\mathcal V$ is twice continuously differentiable on the interior of its domain.
\end{axiom}

These assumptions motivate, but do not uniquely imply, the particular bilinear operational form used below. In particular, nonnegative first-order coefficients are treated as substantive modeling constraints rather than as a consequence of the three axioms alone.

\begin{remark}[Local approximation]
At an interior expansion point $(D_0,M_0)$, a second-order Taylor expansion can include main effects, an interaction term, and higher-order terms. The operational index below should therefore be interpreted as a parsimonious, substantively constrained functional form rather than as the unique function characterized by the axioms.
\end{remark}

\begin{definition}[Operational Vulnerability Index]
Define
\begin{align}
\Disc_S&=\frac{|\dAUC_S|}{c},\\
\Sev_S&=1-\exp(-\lambda\pi_S),
\end{align}
with fixed constants $c$ and $\lambda$, and
\begin{equation}
V(S)=w_1\Disc_S+w_2\Sev_S+w_3\Disc_S\Sev_S.
\end{equation}
\end{definition}

\begin{proposition}[Constrained Optimal Weights]
\label{prop:optimal_weights}
Impose the symmetry constraint $w_1=w_2\equiv t$ (equal treatment of discrimination and severity as main effects), so that $w_3=1-2t$, together with two design constraints motivated by the framework's purpose: (a) \emph{interaction dominance}, $w_3\geq w_1$ and $w_3\geq w_2$, reflecting the requirement that joint occurrence of discrimination loss and severity be weighted at least as heavily as either alone; and (b) a \emph{meaningful main-effect floor}, $t\geq0.25$, ensuring neither main effect vanishes from the index. Fitting $V(S)$ by least squares against the multiplicative target $T=\Disc_S\cdot\Sev_S$ -- the natural ``AND-gate'' benchmark implied by Axiom~\ref{ax:super}, under which vulnerability should require both components to be elevated rather than either alone -- gives the objective
\begin{align*}
\mathcal{L}(t)
&= \EE\big[
(t\,\Disc_S+t\,\Sev_S+(1-2t)\,\Disc_S\Sev_S
-\Disc_S\Sev_S)^2
\big] \\
&= t^2\,\EE\big[
(\Disc_S+\Sev_S-2\,\Disc_S\Sev_S)^2
\big].
\end{align*}

a pure quadratic in $t$ with unconstrained minimizer $t^*=0$. Constraint (a) requires $1-2t\geq t$, i.e. $t\leq1/3$; together with constraint (b) this restricts $t$ to $[0.25,\,1/3]$, an interval lying entirely to the right of the unconstrained minimizer. Since $\mathcal{L}(t)=\kappa t^2$ for the nonnegative constant $\kappa=\EE[(\Disc_S+\Sev_S-2\,\Disc_S\Sev_S)^2]$, which does not depend on $t$, $\mathcal{L}$ is strictly increasing on $(0,\infty)$ and the constrained minimum occurs at the left boundary of the feasible interval:
\[
t^*=0.25 \;\Longrightarrow\; w_1=w_2=0.25,\quad w_3=0.50.
\]
Because $\kappa$ is a fixed nonnegative constant regardless of the joint distribution of $(\Disc_S,\Sev_S)$, this result holds for \emph{any} such distribution -- in particular, it does not depend on the empirical correlation between discrimination and severity across subgroups. The full derivation is given in S1 Appendix, Section 4.
\end{proposition}

The prespecified values were $c=0.30$ and $\lambda=3$; the clinical rationale for $c$ is given in S1 Appendix, Section 1.6. The weights $w_1=w_2=0.25$ and $w_3=0.50$ are not an independent design choice but follow directly from Proposition~\ref{prop:optimal_weights}.

\subsubsection{Asymptotic Properties}
For subgroup $S$, inference for $V(S)$ follows the delta method applied jointly to the subgroup AUC difference and subgroup mortality proportion. Let $\hat\theta_S=(\widehat{\dAUC}_S,\hat\pi_S)^\top$ and let $\hat\Sigma_S$ denote a consistent covariance estimator for $\hat\theta_S$. Under standard regularity conditions and a nonzero population AUC difference,
\begin{equation}
\sqrt{n_S}(\hat\theta_S-\theta_S)\xrightarrow{d}N(0,\Sigma_S).
\end{equation}
Since $V(S)=g(\theta_S)$ with
\begin{equation}
\frac{\partial V}{\partial \dAUC}=\operatorname{sign}(\dAUC_S)\frac{w_1+w_3\Sev_S}{c},
\qquad
\frac{\partial V}{\partial \pi_S}=(w_2+w_3\Disc_S)\lambda e^{-\lambda\pi_S},
\end{equation}
the delta method gives
\begin{equation}
\sqrt{n_S}(\hat V(S)-V(S))\xrightarrow{d}N(0,\tau_S^2),
\end{equation}
where
\begin{equation}
\tau_S^2=\nabla g(\theta_S)^\top\Sigma_S\nabla g(\theta_S).
\end{equation}
The covariance term should be estimated rather than assumed to be zero unless the exact placement-value influence-function conditions used in the implementation are verified. Accordingly, the implementation uses the corresponding joint covariance estimator in $\hat\tau_S^2$. The full derivation, including the regularity conditions under which the asymptotic covariance is diagonal, the closed-form DeLong variance components, and the justification for the normalization constant $c$, is given in S1 Appendix, Section 1.

\subsubsection{Hypothesis Testing and Multiple Comparisons}
The composite target is
\begin{equation}
H_0: V(S)\leq\tau^*\ \text{or}\ |\dAUC_S|\leq\delta_0,
\qquad
H_1: V(S)>\tau^*\ \text{and}\ |\dAUC_S|>\delta_0,
\end{equation}
where $\delta_0=0.05$. The intersection-union principle \citep{berger1982multiparameter} implies that rejection of both component null hypotheses is required to reject $H_0$. For the vulnerability component, a one-sided Wald statistic is used. For the absolute AUC-difference component, the statistic is based on the lower confidence bound for $|\dAUC_S|$; the sign of the estimated difference is retained when constructing the corresponding one-sided local test so that the nondifferentiability at zero is not treated as an interior regular point.

Multiplicity across the $K$ prespecified subgroups is controlled using a patient-level bootstrap with $B=2{,}000$ resamples and a Romano--Wolf stepdown procedure applied to the relevant studentized component statistics. The bootstrap procedure is used to preserve the dependence among subgroup statistics induced by overlapping subgroups. The exact implementation, including the bootstrap studentization, stepdown ordering, and the argument for bootstrap validity given the asymptotic linearity of $\hat V(S)$, is given in S1 Appendix, Section 3, and the code repository.

\subsubsection{Choice of Clinical Constants}
The prespecified constants were $\delta=0.10$ as a minimum clinically meaningful AUC difference, $\delta_0=0.05$ as the composite-rule floor, $\tau^*=0.25$ as the numerical vulnerability threshold, $c=0.30$ as the normalization ceiling, and $\epsilon=0.02$ as the overall-performance tolerance. Rankings were invariant across $c\in\{0.25,0.30,0.40\}$ (Spearman $\rho=1.00$). The clinical rationale for $c$ is given in S1 Appendix, Section 1.6, and the weight derivation in Section 4; justification for the remaining constants ($\delta$, $\delta_0$, $\tau^*$, $\epsilon$) should be documented with the prespecified analysis plan rather than described as being empirically calibrated from the current test results.

\subsubsection{Hierarchical Shrinkage for Small Subgroups}
Partition-wise empirical-Bayes shrinkage via REML was used as a secondary approach for stabilizing estimates in small subgroups. The final estimator was constructed as a reliability-weighted combination of the raw subgroup estimate and its partition-level pooled estimate. Because shrinkage changes both point estimates and uncertainty, the exact estimator, variance estimator, and any small-sample correction are specified in S1 Appendix, Section 2, rather than relying on a generic claim of bias correction.

\subsubsection{Weight Validity Assessment}
Agreement between independent test-split $\hat V(S)$ rankings and validation-split $|\dAUC_S|$ was assessed with Spearman correlation. This was treated as an external consistency check, not as a validation of causal or optimal weighting.

\subsubsection{Aggregate Loss and Equity Penalty}
For each univariate partition, $L(f)=\sum_m\nu(S_m)V(S_m)$ and $\Delta L(f)=\sum_m\nu(S_m)|\dAUC_{S_m}|\mu_{S_m}$, with $\mu_S=w_1+w_3\Sev_S$. The partition-level equity penalty was defined as $\lambda^*=\Delta L(f)/L_{\mathrm{standard}}(f)$, where the denominator was the standard RF Brier baseline. This quantity is interpreted as an index of aggregate vulnerability relative to the stated baseline, not as a causal welfare loss.

\subsubsection{Detection and Sample Size}
The minimum sample-size calculation was based on the desired detectable AUC difference and the variance of the corresponding subgroup estimator. Because AUC variance depends on case-control prevalence and score distributions, the final calculation used the estimated variance components from the proposed AUC estimator rather than a prevalence-only adjustment. The exact numerical implementation should be reported with the assumptions for event prevalence, effect size, and variance in Supporting Information.

\subsubsection{Multi-Model Robustness Extension}
The secondary analysis calculated the maximum observed pairwise $|\dAUC_S|$ among RF, LR, GBM, MLP, and ENet. This broadens the empirical audit but is not treated as a direct estimate of $M(S)$ unless the models entering the maximum satisfy the prespecified $\calF_\epsilon$ criterion.

\subsubsection{Wasserstein-Distance Robustness Check}
An exploratory Wasserstein-distance analysis examined whether subgroup distributional differences were associated with observed model disagreement. No formal generalization bound was assumed. Exact optimal transport was used for the reported empirical check \citep{cuturi2013sinkhorn}; the full transport formulation is given in S1 Appendix, Section 5.1.

\subsubsection{Statistical Analysis}
Analyses used R version 4.5.3. AUC was evaluated using DeLong-type placement-value methods and calibration using the Brier score. Two-sided $\alpha=0.05$ was used for descriptive component analyses, with the composite decision rule using the prespecified one-sided thresholds and multiplicity adjustment described above. Sensitivity analyses varied $t\in\{0.25,0.29,0.33\}$, $\lambda\in\{2,3,5\}$, $c\in\{0.25,0.30,0.40\}$, $\epsilon\in\{0.01,0.02,0.03\}$, and $\tau\in\{0.10,0.15,0.20\}$.

\subsubsection{Simulation Studies}
Study 1 evaluated detection across 30 combinations of subgroup size and true AUC difference, with 1,000 replicates per scenario. Study 2 evaluated prioritization in a planted-failure design. Study 3 compared vulnerability-based and power-inclusive prioritization. Study 4 evaluated empirical Wald-interval coverage across subgroup sizes $10,25,50,100,200,$ and $500$, with 1,000 replicates per level. Complete simulation specifications and results for all four studies are provided in S1 Appendix, Sections 5.2--5.5.

\section{Results}

\subsection{Study Population and Model Performance}
The development cohort comprised $n=65{,}078$ MIMIC-IV admissions; the external-validation cohort comprised $n=188{,}230$ eICU-CRD admissions across 208 hospitals. The pooled validation split contained 50,640 scored patients and the pooled test split contained 50,650 (see Figure~\ref{fig:flow} in Methods for the complete data-flow diagram).

\begin{table}[H]
\centering
\caption{Overall discrimination of candidate models on the validation split.}
\label{tab:candidate_aucs}
\begin{tabular}{lc}
\toprule
Model & AUC\\
\midrule
Random Forest (RF) & 0.8021\\
Logistic Regression (LR) & 0.7533\\
Gradient-Boosted Trees (GBM) & 0.7905\\
Multilayer Perceptron (MLP) & 0.7622\\
Elastic-Net Logistic (ENet) & 0.7538\\
\bottomrule
\end{tabular}
\end{table}

\subsection{Assessment of the Prespecified $\calF_\epsilon$ Tolerance}
RF achieved the best validation AUC among the five candidates, $\mathrm{AUC}^\star=0.8021$. Its gap from the benchmark was therefore 0, whereas the LR gap was 0.0488. Thus LR did not belong to the prespecified $\epsilon=0.02$ comparison class. We retained the prespecified tolerance and did not enlarge it after observing the results. This distinction is important for interpreting the theoretical lower-bound argument: the RF--LR difference remains an observed model-disagreement measure, but the formal inequality linking that difference to the supremum over $\calF_\epsilon$ is not invoked for this application.

\subsection{Prespecified Subgroup Set}
The analysis included $K=158$ prespecified subgroups comprising univariate strata, two-way intersections meeting the $n_S\geq50$ training-data threshold, and clinically motivated three-way intersections.

\subsection{Subgroup Vulnerability Estimates and Composite Decisions}
Nine of 158 subgroups had RF--LR discrimination gaps distinguishable from the prespecified clinical floor. The point estimate for Age $\geq80$ and Cardiac was $\hat V(S)=0.307$, which exceeded the numerical priority threshold $\tau^*=0.25$; however, this subgroup was underpowered and did not satisfy the complete high-priority decision rule. Thus the statement that no subgroup was high priority refers to the full decision rule, not to the absence of point estimates above the numerical threshold. Before and after Romano--Wolf multiplicity adjustment \citep{romano2005exact} ($B=2{,}000$), 0 of 158 subgroups were classified as high priority.

\begin{table}[H]
\centering
\caption{Subgroups with RF--LR discrimination gaps statistically distinguishable from the prespecified clinical floor, ranked by $\hat V(S)$. A point estimate above $\tau^*$ does not by itself constitute a high-priority decision.}
\label{tab:sig_subgroups}
\small
\begin{tabular}{lccccc}
\toprule
Subgroup & Type & $n_S$ & $\hat V(S)$ & 95\% CI & Powered\\
\midrule
Age $\geq$80 \& Cardiac & Two-way & 581 & 0.307 & [0.218, 0.397] & No\\
Age 65--79 \& Cardiac & Two-way & 1,302 & 0.210 & [0.157, 0.264] & No\\
Male \& Cardiac & Two-way & 2,168 & 0.199 & [0.160, 0.238] & No\\
Age 65--79 \& Cardiac \& Male & Three-way & 832 & 0.199 & [0.131, 0.266] & No\\
\textbf{Cardiac (univariate)} & \textbf{Univariate} & \textbf{3,386} & \textbf{0.193} & \textbf{[0.163, 0.223]} & \textbf{Yes}\\
Age 40--64 \& Cardiac & Two-way & 1,389 & 0.189 & [0.134, 0.244] & No\\
Charlson 1--2 \& Cardiac & Two-way & 1,239 & 0.188 & [0.136, 0.240] & No\\
White \& Cardiac & Two-way & 2,395 & 0.174 & [0.140, 0.208] & No\\
Age $<$40 \& Charlson 0 & Two-way & 4,106 & 0.119 & [0.074, 0.165] & No\\
\bottomrule
\end{tabular}
\end{table}

\begin{figure}[H]
\centering
\includegraphics[width=0.9\textwidth]{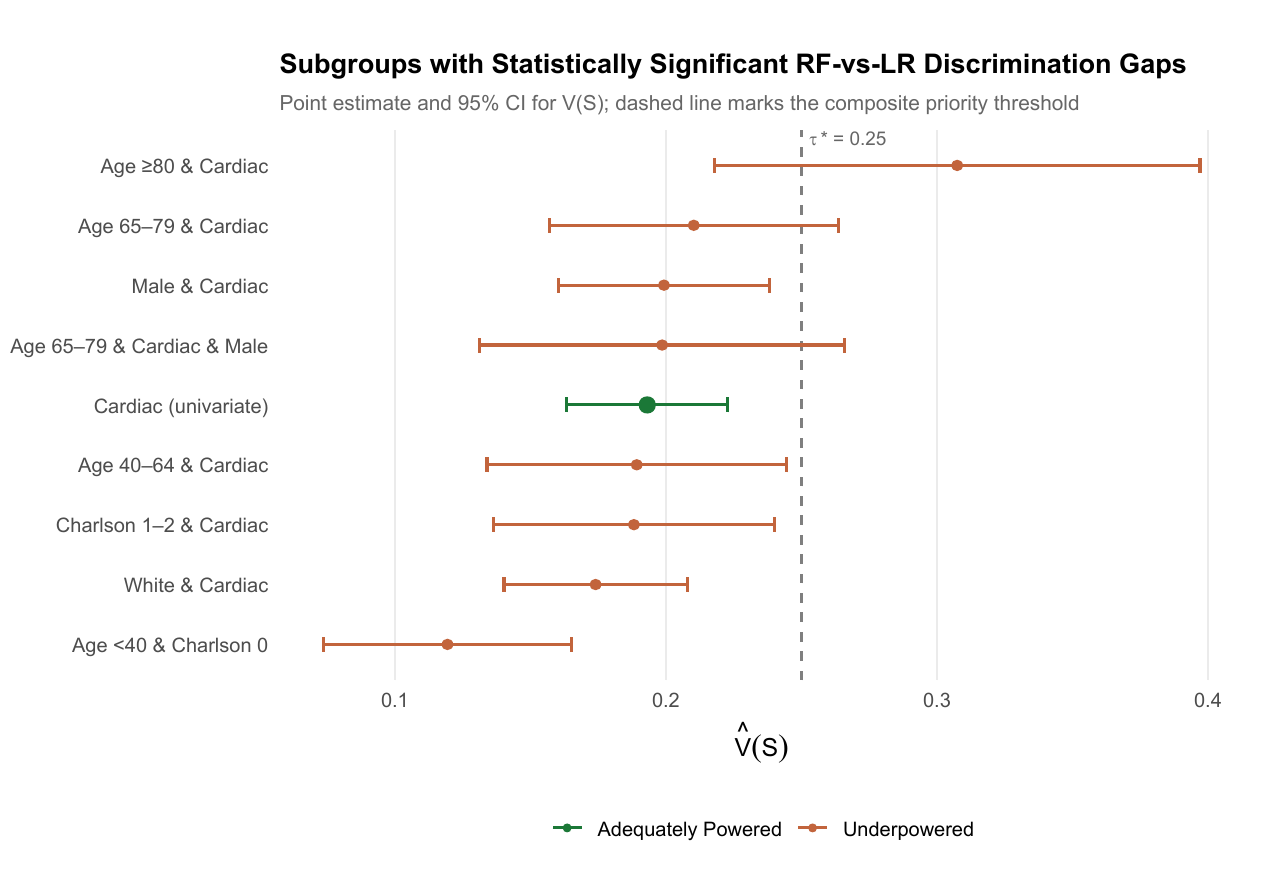}
\caption{\textbf{Subgroups with statistically distinguishable RF--LR discrimination gaps.} Subgroups with $Z_S'>1.96$ are ranked by $\hat V(S)$. Error bars show 95\% Wald confidence intervals. The dashed line marks the numerical priority threshold $\tau^*=0.25$; crossing this line is not sufficient for high-priority classification when the complete composite decision rule is not met.}
\label{fig:forest}
\end{figure}

\subsection{Power Analysis}
Of 156 subgroups with computable $n_{\min}(S)$, 57 (36.5\%) were adequately powered. Power differed sharply by type: 75.0\% of univariate strata, 31.3\% of two-way intersections, and 0.0\% of three-way intersections were adequately powered. The largest point estimate therefore occurred in a subgroup whose statistical uncertainty was substantial.

\subsection{Simulation Validation}
The simulations provide direct evidence about the finite-sample behavior of the proposed procedures before secondary robustness analyses are interpreted.

\textbf{Study 1 (Detection).} Across 30 scenarios ($n_S\in\{25,\ldots,800\}$, true $|\dAUC_S|\in\{0,\ldots,0.20\}$, 1,000 replicates each), empirical detection rates exceeded the corresponding asymptotic predictions at small $n_S$. Under the null at $n_S=25$, empirical detection was 60.6\% versus 29.2\% for the asymptotic approximation. This demonstrates substantial finite-sample miscalibration rather than a minor approximation error. The discrepancy narrowed with sample size, with mean absolute error 0.061 across the 30 scenarios and closer agreement by approximately $n_S\geq200$. Accordingly, nominal error-rate interpretations should not be extended to very small subgroups without additional finite-sample calibration.

\textbf{Study 2 (Prioritization).} In a planted-failure design, $V(S)$-based ranking identified the planted subgroup as highest priority in 74.6\% of replicates, compared with 11.4\% for $|\dAUC_S|$-only ranking.

\textbf{Study 3 (Power Exclusion).} In a design contrasting a small severe subgroup with a large trivial subgroup, $V(S)$ correctly prioritized the small severe subgroup in 100.0\% of replicates, compared with 94.0\% for the power-inclusive alternative. This supports the conceptual distinction between vulnerability and statistical detectability, but does not imply that the vulnerability index is well calibrated at every subgroup size.

\textbf{Study 4 (CI Coverage).} Empirical coverage of the Wald intervals ranged from 91.4\% at $n_S=10$ to 95.6\% at $n_S=200$. Coverage approached the nominal 95\% level by approximately $n_S\geq100$, whereas smaller subgroups showed measurable undercoverage. We therefore treat Wald inference for very small subgroups as approximate and use the simulation results to motivate shrinkage and/or finite-sample calibration rather than claiming uniform validity across subgroup sizes.

\begin{table}[H]
\centering
\caption{Simulation Study 4: empirical coverage of asymptotic Wald confidence intervals.}
\label{tab:study4}
\begin{tabular}{lccc}
\toprule
$n_S$ & Empirical Coverage & Nominal & Nominal $-$ Empirical\\
\midrule
10 & 0.914 & 0.95 & 0.036\\
25 & 0.929 & 0.95 & 0.021\\
50 & 0.945 & 0.95 & 0.005\\
100 & 0.955 & 0.95 & $-$0.005\\
200 & 0.956 & 0.95 & $-$0.006\\
500 & 0.943 & 0.95 & 0.007\\
\bottomrule
\end{tabular}
\end{table}

\subsection{Multi-Model Robustness Check}
Sixty-seven of 158 subgroups (42.4\%) showed at least one pair among the five candidate models exceeding the clinical floor, compared with 9 of 158 (5.7\%) for RF versus LR. RF contributed to the majority of maximal disagreements, most often against ENet or LR. Because LR was outside the prespecified $\epsilon$-class, this analysis is presented as a robustness and sensitivity analysis rather than as a direct empirical evaluation of the theoretical supremum over $\calF_\epsilon$.

\subsection{Weight Validity Assessment}
The Spearman correlation between test-split $\hat V(S)$ rankings and independent validation-split $|\dAUC_S|$ was $\rho=0.032$. This is essentially no rank agreement and provides little independent evidence that the particular subgroup ordering is stable across splits. It should therefore be interpreted as a limitation of ranking stability, not as evidence supporting the proposed weights (the weight derivation is detailed in S1 Appendix, Section 4).

\subsection{Supplementary Wasserstein-Distance Check}
The Spearman correlation between $d_W(p_T,p_S)$ and $|\dAUC_S|$ was $\rho=-0.138$ ($n=156$ subgroups, $n_S\geq25$), indicating a weak association. This analysis is exploratory and does not establish a generalization relationship between distributional distance and predictive multiplicity (full specification in S1 Appendix, Section 5).

\subsection{Aggregate Deployment Loss and Equity Penalty}
\begin{table}[H]
\centering
\caption{Aggregate deployment loss and equity penalty by partition.}
\label{tab:aggregate_loss}
\begin{tabular}{lccc}
\toprule
Partition & $L(f)$ & $\Delta L(f)$ & Equity Penalty\\
\midrule
Age group & 0.1205 & 0.0187 & 0.2644\\
Sex & 0.1136 & 0.0160 & 0.2263\\
Race/ethnicity & 0.1093 & 0.0150 & 0.2120\\
Charlson category & 0.1154 & 0.0170 & 0.2398\\
Diagnosis category & 0.1117 & 0.0156 & 0.2210\\
\bottomrule
\end{tabular}
\end{table}
The maximum partition-level penalty was $\lambda^*=0.2644$ for age group.

\subsection{Cross-Site Generalizability}
145 of 208 eICU-CRD sites met the $n\geq50$ test-split threshold. Across these sites, estimated $\hat V(S)$ values ranged from 0.032 to 0.394. This heterogeneity indicates that subgroup vulnerability may vary across hospitals and supports site-specific auditing in multi-institutional deployments.

\subsection{Equity-Penalized Model}
The equity-weighted Random Forest had an overall Brier score of 0.0742, compared with 0.0708 for the standard Random Forest. This represents a modest aggregate calibration cost; the present analysis does not establish the corresponding subgroup-specific discrimination benefit.

\subsection{Mechanistic Characterization of the Cardiac-Diagnosis Subgroup}
Within the cardiac subgroup ($n=3{,}386$), the highest predicted-risk decile had mean predicted mortality of 0.446 and observed mortality of 0.543 for RF, compared with 0.251 and 0.404 for LR. Permutation importance ranked lactate highly for RF in this subgroup (0.0259 versus 0.0087 in non-cardiac subgroups), while lactate also had the largest reported LR coefficient ($\hat\beta=0.484$, $z=12.17$, $p=4.6\times10^{-34}$). These findings show that lactate is strongly used by both models, but they do not by themselves establish that differences in lactate functional form caused the observed RF--LR disagreement. The lactate analysis is therefore treated as post hoc characterization rather than formal mediation or mechanistic identification.

\begin{table}[H]
\centering
\caption{Calibration by predicted-risk decile in the cardiac subgroup.}
\label{tab:calibration_cardiac}
\begin{tabular}{lcc}
\toprule
 & RF (decile 10) & LR (decile 10)\\
\midrule
Mean predicted risk & 0.446 & 0.251\\
Observed mortality & 0.543 & 0.404\\
$n$ & 339 & 339\\
\bottomrule
\end{tabular}
\end{table}

\begin{figure}[H]
\centering
\includegraphics[width=0.75\textwidth]{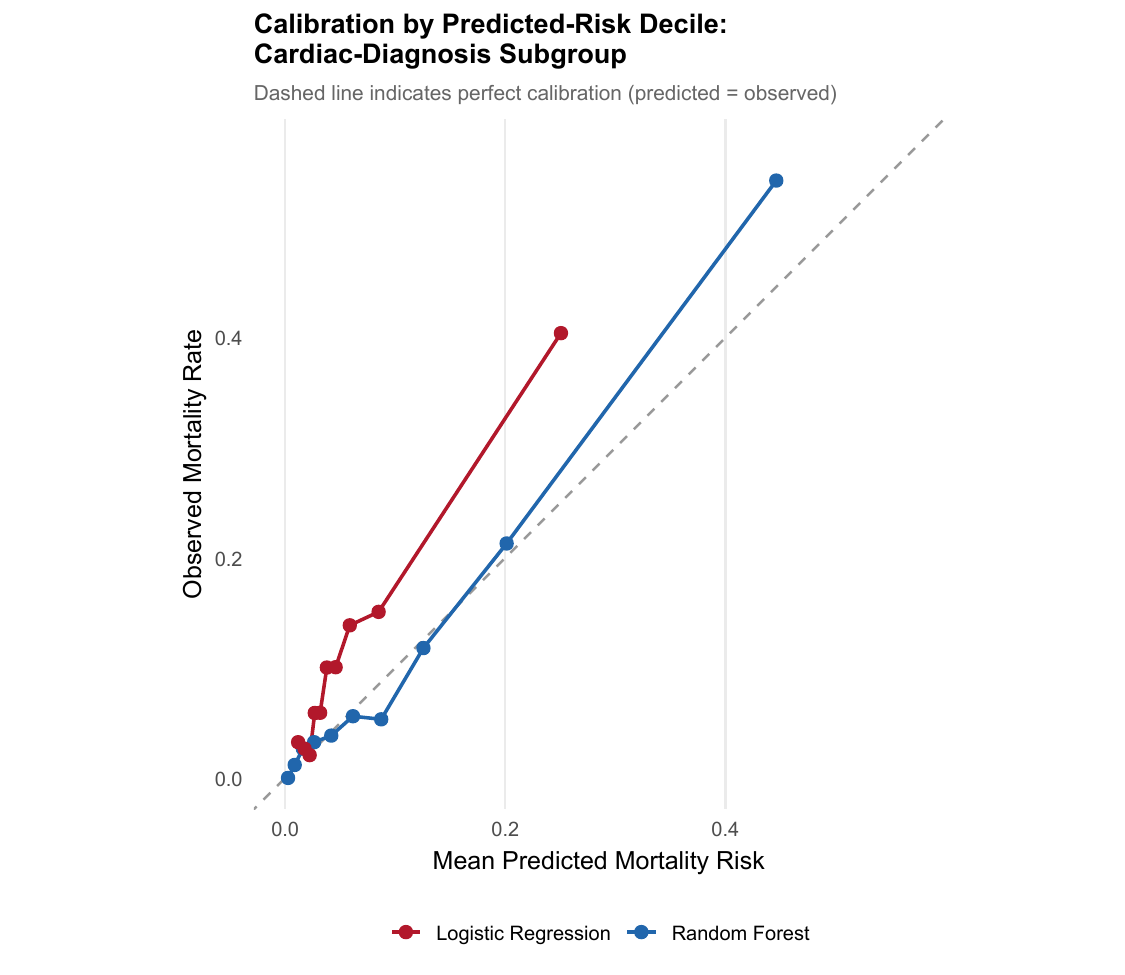}
\caption{\textbf{Predicted and observed mortality in the highest-risk decile of the cardiac subgroup, RF versus LR.} The figure is descriptive and is not intended to establish a causal mechanism for model disagreement.}
\label{fig:calibration}
\end{figure}

\section{Discussion}

\subsection{Principal Findings}
The framework identified the univariate cardiac-diagnosis subgroup as the only adequately powered subgroup with a discrimination gap distinguishable from the prespecified clinical floor. Several smaller intersections involving cardiac diagnosis showed similar directional signals, but their limited power and small-sample inferential limitations make them less secure findings. The Age $\geq80$ and Cardiac subgroup illustrates this distinction particularly clearly: its point estimate of $V(S)=0.307$ exceeded the numerical threshold, yet its confidence interval was wide and it did not meet the complete high-priority decision rule. Thus the framework separates a large observed point estimate from a sufficiently supported deployment-priority decision.

No subgroup met the complete high-priority threshold after multiplicity adjustment. This should not be interpreted as evidence that no clinically meaningful disagreement exists. Rather, under the prespecified decision rule, the available evidence was insufficient to classify any subgroup as high priority while controlling multiplicity. The nine discrimination-gap signals and their uncertainty intervals remain useful for prioritizing future validation.

\subsection{Interpretation of the Rashomon-Set Premise}
The theoretical lower-bound result is conditional on both compared models belonging to $\calF_\epsilon$. In this application, RF satisfied the criterion but LR did not at $\epsilon=0.02$. We therefore do not claim that the observed RF--LR gap is a formal lower bound on disagreement over the full $\epsilon$-Rashomon set. Instead, the primary application should be understood as an audit of disagreement between two specific models selected a priori for their complementary modeling characteristics. The broader five-model analysis demonstrates that comparator choice matters, but it also cannot recover a formal Rashomon-set guarantee unless the candidate models satisfy the prespecified overall-performance criterion.

\subsection{Power and Small-Sample Inference}
Power decreased sharply with subgroup specificity. This is expected because intersections rapidly reduce the number of outcome events and non-events available for estimating subgroup-specific discrimination. The simulation results add an important qualification: the asymptotic composite procedure is poorly calibrated at very small subgroup sizes, and Wald intervals show measurable undercoverage. These findings mean that a statistically significant result in a small subgroup should not be interpreted solely through its nominal asymptotic $p$-value. Larger validation cohorts, shrinkage, bootstrap calibration, or alternative finite-sample procedures are preferable when auditing sparse intersections.

\subsection{Multi-Model Disagreement}
The increase from 5.7\% of subgroups with an RF--LR gap to 42.4\% with at least one pair among five candidate models demonstrates substantial dependence of the observed signal on comparator choice. This is consistent with the conceptual motivation of predictive multiplicity, but because not all five candidates lie in the prespecified $\epsilon$-class, the multi-model result should be regarded as an empirical robustness analysis rather than a direct estimate of the theoretical supremum $M(S)$.

\subsection{Ranking Stability and Site Heterogeneity}
The near-zero validation/test rank correlation ($\rho=0.032$) is an important limitation. It suggests that the ordering of subgroup vulnerability estimates is sensitive to the data split, even though individual subgroup estimates can still be informative. The site-level range of 0.032--0.394 further indicates heterogeneity across hospitals. Together, these results argue for treating subgroup rankings as candidates for validation rather than immutable rankings of clinical risk.

\subsection{Lactate Characterization}
The cardiac-subgroup analysis suggests a plausible area for further investigation because lactate was important to both models and the models differed substantially in predicted risk among high-risk cardiac patients. However, the available analyses do not identify the functional form learned by RF or establish that lactate alone generated the AUC difference. The finding should therefore be considered hypothesis-generating. Formal partial-dependence, interaction, calibration, or counterfactual analyses would be needed to establish a more specific explanation.

\subsection{Limitations}
Several limitations should guide interpretation. First, the primary RF--LR comparison does not satisfy the prespecified $\epsilon$-Rashomon membership condition for LR, so the formal lower-bound interpretation is not available in the application. Second, the proposed asymptotic inference is not reliable at very small subgroup sizes, as demonstrated by the simulation results. Third, the test/validation ranking correlation was essentially zero, limiting evidence for stable subgroup ordering. Fourth, the Wasserstein analysis is exploratory and showed only a weak association. Fifth, the equity-weighted RF incurred a modest Brier-score cost, but the present analysis does not quantify whether that cost yields a clinically meaningful subgroup benefit. Sixth, comorbidity burden was not measured on a common scale across data sources: MIMIC-IV uses a weighted Charlson comorbidity index derived from ICD-9/10 diagnosis codes, whereas eICU-CRD uses an unweighted count of eight comorbidity flags available in \texttt{apachePredVar} as a proxy, given the absence of comparably granular diagnosis coding. Both were mapped to the same categorical strata (0, 1--2, 3+) for subgroup definition, so any Charlson-stratified subgroup (e.g., ``Charlson 1--2 \& Cardiac'') combines two non-equivalent underlying measures rather than a harmonized comorbidity burden. Finally, race/ethnicity harmonization across MIMIC-IV and eICU-CRD required collapsing categories, leaving potential residual misclassification. In addition, missingness in lactate and temperature was concentrated rather than random -- consistent with these labs and vitals being ordered selectively based on clinical suspicion rather than routinely -- so training-split median imputation may not fully remove associated bias, particularly if the missingness mechanism differs across the subgroups examined.

\subsection{Practical Implementation}
In practice, $V(S)$ is a pre-deployment and periodic post-deployment audit for model governance teams, not a point-of-care tool. The audit runs quickly on a standard workstation and requires only a labeled validation dataset---already collected for routine performance monitoring---and the analysis code accompanying this paper, archived at Zenodo (\url{https://doi.org/10.5281/zenodo.22339280}; see Data and Code Availability). Developers would compute $V(S)$ across prespecified subgroups, apply the composite decision rule with multiplicity correction, and route high-priority subgroups to targeted validation rather than aggregate review alone. The cross-site results reported above ($V(S)$ ranging from 0.032 to 0.394 across 145 hospitals) indicate the audit should be repeated at each deployment site rather than performed once centrally. A high-priority subgroup should trigger one of three actions: collect additional data, apply hierarchical shrinkage as an interim adjustment (S1 Appendix, Section 2), or restrict model use pending further evaluation---with the choice left to institutional risk tolerance.

\section{Conclusion}
This study develops a statistical framework for identifying patient subgroups in which clinical prediction models may disagree in ways that warrant additional validation. In the application, the broad cardiac-diagnosis subgroup provided the strongest adequately powered signal, while smaller cardiac intersections produced less certain estimates. The analysis also demonstrates why subgroup auditing should report model-comparison assumptions, statistical power, multiplicity adjustment, and finite-sample operating characteristics rather than relying on point estimates alone. The framework is best viewed as a prospective auditing and prioritization tool: it can identify where further investigation is warranted, but it does not by itself establish that one model is biased, unsafe, or causally responsible for a subgroup disparity.

\section*{Acknowledgments}
The author thanks Dr. Edward Bedrick and Gongzhong Yao for their contributions to this work.

\section*{Author Contributions}
Enock A.B. conceived the study, developed the theoretical framework, performed the data extraction, harmonization, analysis, and simulations, and wrote the manuscript.

\section*{Funding Statement}
This research received no specific grant from any funding agency in the public, commercial, or not-for-profit sectors.

\section*{Competing Interests}
The author has declared that no competing interests exist.

\bibliographystyle{unsrtnat}
\bibliography{References}

\newpage
\section*{Supporting Information}

\noindent\textbf{S1 Appendix.} Asymptotic inference, weight derivation, hierarchical shrinkage, and simulation specifications for the vulnerability index. Section 1 gives the joint influence-function representation, covariance estimator, delta-method derivation, closed-form DeLong variance, and justification for the discrimination normalization constant $c$. Section 2 gives the partition-wise REML shrinkage model, reliability weights, and parametric-bootstrap posterior variance. Section 3 gives the argument for bootstrap validity underlying the Romano--Wolf multiplicity correction. Section 4 gives the complete constrained-MSE derivation of the weights $w_1=w_2=0.25$, $w_3=0.50$ (Proposition~\ref{prop:optimal_weights}). Section 5 gives the Wasserstein-distance transport formulation and the complete specifications and results of Simulation Studies 1--4.

\end{document}